\documentclass[twocolumn]{jpsj3owmc}

\usepackage{txfonts}
\usepackage{color}

\title{Superconductivity and inhomogeneous charge-ordered state in the
two-Dimensional Hubbard model\\
-- Off-Diagonal Wave Function Monte Carlo Studies of Hubbard Model IV --
}

\author{Takashi Yanagisawa$^{1,2,*}$}

\inst{
$^1$Electronics and Photonics Research Institute, 
Core Manufacturing Technology Research Institute,
Department of Electronics and Manufacturing,
National Institute of Advanced Industrial Science and Technology (AIST),
1-1-1 Umezono, Tsukuba, Ibaraki 305-8568, Japan\\
$^2$Faculty of Health Science, Tsukuba University of Technology, 
4-12-7 Kasuga, Tsukuba, Ibaraki 305-8521, Japan
}

\abst{We investigate the ground-state properties of the
two-dimensional Hubbard model, based on the off-diagonal wave function variational 
Monte Carlo method.
We use an optimized wave function that is improved from an initial one-body wave function 
by multiplying by multiple correlation operators that are given by the form
of exp($-S$)--type where $S$ is a suitable operator.
We examine the inhomogeneous ground state at near 1/8 doping in the strongly correlated 
region where the on-site Coulomb interaction is larger than the bandwidth.  
The $d$-wave superconductivity with a spatially oscillating gap function can coexist with
the charge ordering in the ground state without magnetic ordering, and superconducting 
condensation energy increases by this coexistence.
We also show the $d$--wave pair correlation function as a function of lattice sites.
The correlation function indicates that the long-range superconducting order indeed exists,
and also that the strong electron correlation suppresses the pair correlation function.
}
 
\begin{document}
\maketitle

\section{Introduction}

The physics and mechanism of high-temperature superconductors have been studied intensively
since the discovery of high-temperature superconductivity\cite{bed86}.   
The parent materials of high-temperature cuprates are Mott insulators when no
carriers are doped, and thus high-temperature cuprates are typical strongly
correlated electron systems.
It is primarily important to understand the electronic properties of the ground state in
strongly correlated electron systems.

The CuO$_2$ plane that is commonly contained in high-temperature cuprates
consists of oxygen atoms and copper atoms, and plays an important role in the
emergence of high temperature superconductivity\cite{mce03,hus03,web08,hyb89,esk89,mcm90,esk91}.
The fundamental electronic model for this plane is given by the d-p model 
(or three-band Hubbard
model)\cite{web08,hyb89,esk89,mcm90,esk91,eme87,hir89,sca91,ogu94,koi00,yan01,koi01,yan03,koi03,
koi06,yan09,web09,lau11,web14,ave13,ebr16,tam16,yan21d}.
The two-dimensional (2D) single-band Hubbard model\cite{hub63,hub64,gut63} is also
the basic model for cuprates since this model is regarded as a simplified effective model of 
the three-band d-p model.
The 2D Hubbard model contains the important physics that may induce electron pairings
leading to high-temperature superconductivity.
Thus the 2D Hubbard model has been investigated to clarify the mechanism of superconductivity 
in cuprates high-temperature 
superconductors\cite{zha97,zha97b,yan96,nak97,yam98,yam00,yam11,
har09,yan13a,bul02,yok04,yok06,yok13,miy02,yan08,yan13,yan98,yan16,yan19,yan19b,dar18,qin20}.
Concerning the origin of the effective attractive interaction from the on-site Coulomb
repulsive interaction, the ladder Hubbard models\cite{noa95,noa97,yam94,yan95,koi99,nak07,jia20} 
have also been studied.

The phase diagram of the 2D Hubbard model has been examined intensively and recent
studies pointed out the existence of superconducting (SC) phase in the ground 
state\cite{nak97,yam98,yok04,yok13,yan16,dar18}.
The next nearest-neighbor transfer integral $t'$ plays an essential role in determining the
phase diagram of magnetic and SC phases in the ground state.  
There are three important parameters; they are 
the parameters $t'$, the Coulomb repulsion $U$ and the carrier density.
These three parameters provide rich structures of the phase diagram including superconducting
magnetic phases.
In the case of $t'=0$, the antiferromagnetic (AF) correlation is suppressed when 
holes (carriers) are doped.
There are three fundamental phases in the phase diagram: 
the antiferromagnetic insulator phase (AFI), the coexistence state of
antiferromagnetism and superconductivity and the $d$-wave superconducting state.
The phase diagram for $t'=0$ is described as follows.
When $t'=0$ and $U/t$ is as large as the bandwidth or greater than the bandwidth, 
near half-filling for approximately $0<x \lesssim 0.06$, the ground state is an AF insulator,
where $x$ is the doping rate (carrier density).\cite{yan19} 
The AF insulator phase exists because of an instability toward the phase-separated
state when $t'=0$.
We have the coexistent state for $0.06\lesssim x \lesssim 0.09$.
When the doping rate is the range $x>0.09$, the ground state is pure $d$-wave SC state.
The phase diagram indicates that high-temperature superconductivity may occur
in the strongly correlated region of the Hubbard model.

The existence of inhomogeneous states such as
stripes\cite{tra96,suz98,yama98,ara99,moc00,wak00,bia96,mai10,mon12,bia13,yam21,miy22,yan01b,yin14,yan21c} 
and checkerboard-like density wave states\cite{hof02,wis08,han04,miy09} is also important
in the study of correlated electron systems, in particular,
in high-temperature cuprates.  
The physics of inhomogeneous electronic states has been investigated and their
properties can be understood by using the 2D Hubbard model. 
Recently the coexistence of stripes and superconductivity has been studied intensively
for the ladder Hubbard model.\cite{jia22,xu22,xu24}

We employ the off-diagonal wave function variational Monte Carlo method that is a 
reliable and useful 
tool to investigate strongly correlated electron systems where
we calculate the expectation values numerically by using a Monte Carlo
method\cite{nak97,yam98,yam00,yam11,har09,yan13a}.
A variational wave function can be improved by introducing new correlation operators.
In this paper we introduce correlation operators of $\exp(-S)$-type where $S$ is a
correlation operator\cite{yan98,yan16}.
The Gutzwiller function is written in this form and
an optimization procedure can be performed in a systematic way by
multiplying by the exponential-type operators repeatedly\cite{yan98}.
The ground-state energy is indeed lowered considerably by using
this type of wave function\cite{yan16}.

We have studied the 2D Hubbard model by employing the exponential-type wave
function.  We briefly explain the papers numbered I to III\cite{yan98,yan16,yan19}.
In the paper I, we proposed many-body wave function multiplied by exponential
operators for the 2D Hubbard model.  We showed that  the wave function can be improved and
optimized by multiplying by correlation operators repeatedly. 
In the paper II, we showed that $d$-wave superconducting phase exists in the
strongly correlated region where the on-site Coulomb repulsive interaction is 
as large as the bandwidth or more than the bandwidth.  The antiferromagnetic correlation
is in general very strong when $U$ is large in the 2D Hubbard model.
It was found that the strong antiferromagnetic correlation is suppressed by
doped hole carriers when $U$ is as large as the bandwidth.  In this region
the pure $d$-wave superconducting phase exists. 
In the paper III, we investigated the phase diagram as a function of the doping
rate and we showed that an insulating antiferromagnetic phase exists
near the half-filled case in the strongly correlated region with vanishingly small 
next nearest-neighbor transfer $t'$.
 
After the paper III, we have examined the region including the optimal doping rate
by employing the off-diagonal wave function Monte Carlo method.
The present paper shows the results after III and is organized as follows.  
In Section 2 we show the method of
off-diagonal wave function variational Monte Carlo simulation.  In Section 3, we discuss 
the ground-state properties of the correlated electron state.  The Section 4 is
devoted to the examination of inhomogeneous charge-ordered state and its
coexistence with superconductivity.  In Section 5, we evaluate the $d$-wave pair
correlation function and the SC order parameter.
We give a summary in Section 6.

\section{Method and many-body correlated wave functions}

\subsection{Hamiltonian}

The two-dimensional Hubbard model is written as
\begin{equation}
H= \sum_{ij\sigma}t_{ij}c^{\dag}_{i\sigma}c_{j\sigma}
+U\sum_in_{i\uparrow}n_{i\downarrow},
\end{equation}
where $t_{ij}$ indicates the transfer integral and $U$ is the strength
of the on-site Coulomb interaction.
We set $t_{ij}=-t$ when $i$ and $j$ are nearest-neighbor pairs $\langle ij\rangle$
and $t_{ij}=-t'$ when $i$ and $j$ are next-nearest-neighbor pairs.
$N$ and $N_e$ denote the number of lattice sites and the number of electrons, respectively.
The energy unit is given by $t$.

\subsection{Improved correlated wave functions}

We use the off-diagonal wave function variational Monte Carlo method where the wave 
function contains
strongly correlated effect in its structure.
The wave function is constructed from the Gutzwiller function
\begin{equation}
\psi_G = P_G\psi_0,
\end{equation}
where $P_G$ stands for the Gutzwiller operator 
$P_G=\prod_i(1-(1-g)n_{i\uparrow}n_{i\downarrow})$
with the variational parameter in the range of $0\le g\le 1$.
$\psi_0$ is a suitable one-body wave function.

The wave function is given by\cite{yan98,yan16,yan19,bae87,ots92,eic07,bae09,bae11}
\begin{equation}
\psi_{\lambda}= e^{-\lambda K}\psi_G,
\end{equation}
where $K$ is the non-interacting part of the Hamiltonian:
\begin{equation}
K= \sum_{ij\sigma}t_{ij}c^{\dag}_{i\sigma}c_{j\sigma}.
\end{equation}
$\lambda$ is a real variational parameter which is determined to minimize the 
ground-state energy.
The expectation value of physical quantities can be calculated by using the
auxiliary-field method in Monte Carlo simulations.\cite{yan98,yan07}

\subsection{Correlated superconducting wave function}

For the correlated superconducting state, we take $\psi_0$ as the BCS wave function:
\begin{equation}
\psi_{BCS}= \prod_k (u_k+v_k c_{k\uparrow}^{\dag}c_{-k\downarrow}^{\dag})|0\rangle,
\end{equation}
where $u_k$ and $v_k$ are coefficients appearing in the ratio
$u_k/v_k= \Delta_k/(\xi_k+\sqrt{\xi_k^2+\Delta_k^2})$, where $\Delta_k$ is the gap
function and $\xi_k= \epsilon_k-\mu$ is the dispersion relation.
We use the $d$-wave symmetric gap function $\Delta_k= \Delta_{s}(\cos k_x-\cos k_y)$
with a variational parameter $\Delta_{s}$.

In the evaluation of the correlated superconducting wave function
\begin{equation}
\psi_{\lambda-BCS}= e^{-\lambda K}P_G\psi_{BCS},
\end{equation}
we use the electron-hole transformation for the down-spin electrons\cite{yok88,yan99}:
$d_k= c_{-k\downarrow}^{\dag}$, $d_k^{\dag}= c_{-k\downarrow}$, and the operators for
up-spin electrons remain the same: $c_k= c_{k\uparrow}$.
In the real space this corresponds to $c_i= c_{i\uparrow}$ and $d_i= c_{i\downarrow}^{\dag}$.

\subsection{Antiferromagnetic, striped and nematic states}

We also consider the state with magnetic and charge orders.  The initial state is given by
the eigenstate of the following one-particle Hamiltonian,
\begin{equation}
H_{stripe}= \sum_{ij\sigma}t_{ij}c^{\dag}_{i\sigma}c_{j\sigma}+
\sum_{i\sigma}(\rho_i-{\rm sgn}(\sigma)m_i)n_{i\sigma},
\end{equation}
where $\rho_i$ and $m_i$ stand for the charge and spin modulations, respectively.
They are described as
\begin{align}
\rho_i &= \rho\cos({\bf Q}_c\cdot ({\bf r}_i-{\bf r}_0)),\\
m_i &= m\sin({\bf Q}_{s}\cdot ({\bf r}_i-{\bf r}_0)),
\end{align}
where $\rho$ and $m\equiv \Delta_{AF}$ are regarded as variational parameters.
${\bf r}_0$ denotes the position of the domain boundary.
Two incommensurate wave vectors ${\bf Q}_c$ and ${\bf Q}_s$ characterize the charge and
spin configurations, respectively.
For the commensurate antiferromagnetic (AF) state, we take ${\bf Q}_s={\bf Q}_{AF}=(\pi,\pi)$
and ${\bf Q}_c=(0,0)$ with $\rho=0$.  The stripe state is described by incommensurate
wave vectors ${\bf Q}_c$ and ${\bf Q}_{s}$.
The vertical stripe is represented by ${\bf Q}_s={\bf Q}_{VS}=(\pi\pm 2\pi\delta,\pi)$ where 
$\delta$ indicates the incommensurability defined as the inverse of the period of the
antiferromagnetic order in the $x$-direction.
This indicates the state with two adjacent AF magnetic domains separated by a one-dimensional 
domain
wall in the $y$-direction.  There is a $\pi$-phase shift between two domain walls.
The charge modulation period is given by the half of that of the spin modulation
where we set ${\bf Q}_c= 2{\bf Q}_s$. 

The diagonal stripe state is represented by 
${\bf Q}_c={\bf Q}_{DS}=(\pi\pm 2\pi\delta, \pi\pm 2\pi\delta)$, where the domain wall appears
in the diagonal direction on the lattice.
We consider the site-centered domain boundary where we set ${\bf r}_0=(0,0)$.

We also examine the state with charge order and without magnetic order.
In this paper we call this state the nematic state, that is, the wave function
with $\Delta_{AF}=0$ and $\rho\neq 0$.

\subsection{Method of optimization for optimal variational parameters}

In the process of searching optimal values of variational parameters that minimize
the ground-state energy, we can employ the simultaneous measurements method or
correlated measurements method.\cite{umr88}
In one Monte Carlo run, we can evaluate expectation values for several values of
the variational parameters simultaneously.  This gives an efficient way to find the
minimum of the ground-state energy in the space of multiple parameters since the
Monte Carlo statistical errors are common for a set of parameters in the same Monte Carlo
run.  We shift the variational parameters by small amounts and calculate the expectation
values to find the direction in the parameter space along which the ground-state energy 
decreases.

We use the periodic boundary condition in one direction and the antiperiodic
boundary condition in the other direction.
We use the Metropolis algorithm in the evaluation of expectation values and
we have about $10^6$ steps in one Monte Carlo run.

\section{Properties of correlated states}

\subsection{Momentum distribution}

We examine the two-dimensional Hubbard model on a $16\times 16$ lattice.
We have two parameters $g$ and $\lambda$ for the wave function $\psi_{\lambda}$.
We show the ground-state energy $E$ as a function of $g$ for several values of
$\lambda$ for $U=18t$, $t'=0$ and the number of electrons $N_e=228$ in Fig. 1.
The dependence of $E$ on $\lambda$ is shown in Fig. 2 where we set $g=0.055$.
The ground-state energy changes rapidly and its minimum is at the narrow bottom
as a function of $\lambda$.
This shows that the inclusion of the parameter $\lambda$ is very effective in
lowering the ground-state energy.

\begin{figure}
\centering
\includegraphics[width=7.5cm]{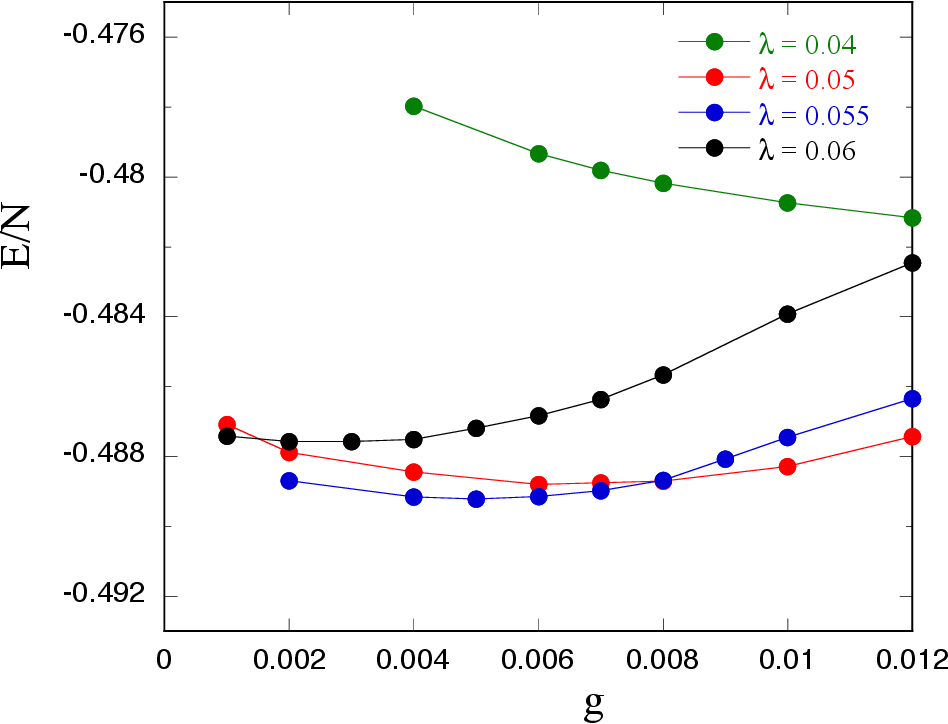}
\caption{(Color online)
The ground-state energy per site $E/N$ as a function of $g$ for $\lambda=0.04$, 
0.05, 0.055 and 0.06, where we put $U=18t$, $t'=0$ and the number of electrons
is $N_e=228$ on a $16\times 16$ lattice.
The Monte Carlo statistical errors are within the size of symbols.
}
\label{fig1}
\end{figure}

\begin{figure}
\centering
\includegraphics[width=7.0cm]{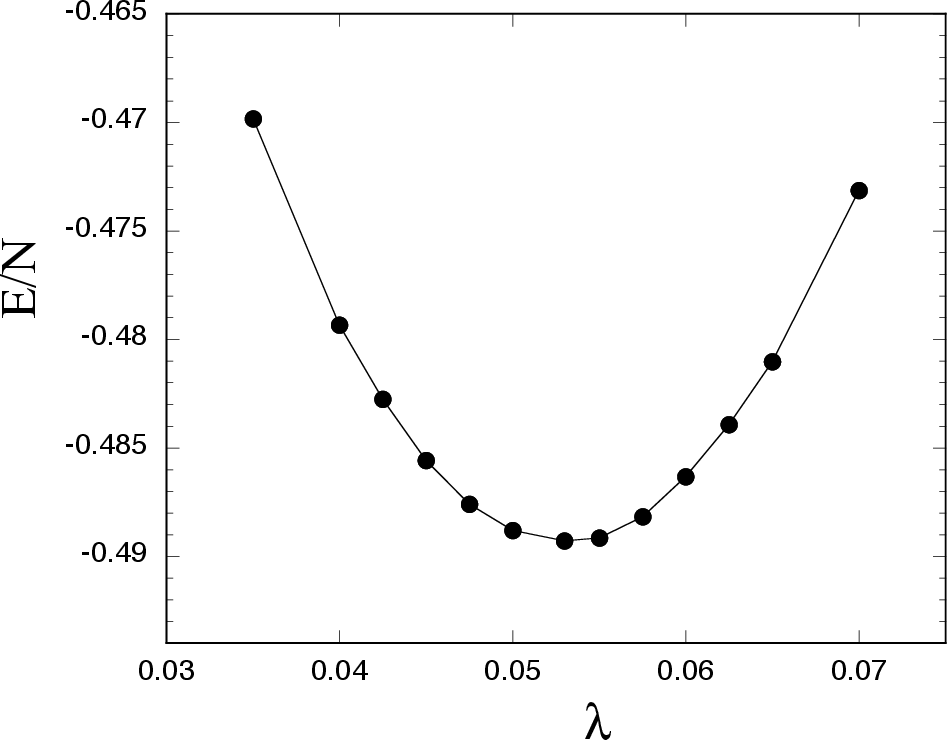}
\caption{
The ground-state energy per site $E/N$ as a function of of $\lambda$  for $g=0.055$,
where $N_e=228$, $U=18t$ and $t'=0$ on a $16\times 16$ lattice.
The Monte Carlo statistical errors are within the size of symbols.
}
\label{fig2}
\end{figure}
 
This state is a highly correlated state with strong on-site repulsive interaction.
This is shown by the momentum distribution function $n({\bf k})$ which is shown in Fig. 3
for a set of parameters indicated above, where the momentum distribution is define as
\begin{equation}
n({\bf k})= \frac{1}{2N}\sum_{iji\sigma}e^{i{\bf k}\cdot ({\bf R}_i-{\bf R}_j)}
\langle c_{i\sigma}^{\dag}c_{j\sigma}\rangle.
\end{equation}
The function $n({\bf k})$ indicates the reduced jump at the Fermi wave number
as a result of the strong correlation.
This figure can be compared to the state with smaller $U$ such as $U=10t$.
In Fig. 3 we also show $n({\bf k})$ for $U=10t$ where band
parameters are $N_e=228$ and $t'=0$ on the $16\times 16$ lattice.
The jump of $n({\bf k})$ at the Fermi wave number becomes larger for smaller $U$.

\begin{figure}
\centering
\includegraphics[width=7.5cm]{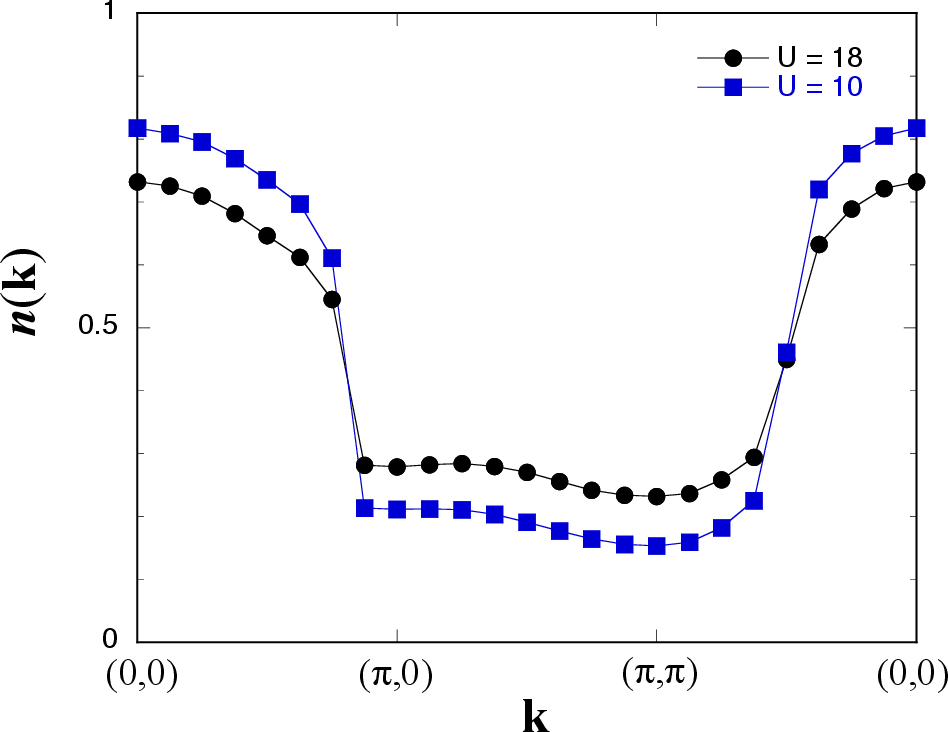}
\caption{(Color online)
The momentum distribution function for $N_e=228$, $U=18t$ and $U=10t$ with $t'=0$ on a
$16\times 16$ lattice.
}
\label{fig3}
\end{figure}

\subsection{Superconductivity}

The existence of superconducting phase in the 2D Hubbard model has been
investigated by using various methods of calculations.
In the off-diagonal wave function variational Monte Carlo method with improved wave
functions, the superconducting phase exists in the strongly correlated region
with large Coulomb repulsion $U$.
In our previous papers we examined the phase diagram including superconductivity
on a $10\times 10$ lattice before\cite{yan16,yan19}.  We study the stability of 
superconducting state
on a $16\times 16$ lattice in this paper.  The system size is larger than that
employed in previous studies.

In the real space, the $d$-wave anisotropic SC order parameters are assigned for each bond 
between the site $i$ and its nearest neighbor site $i+\hat{\mu}$ where $\mu=x,y$ and 
$\hat{\mu}$ indicates the unit vector in the $\mu$-th direction, for which we denote the SC 
order parameter as $\Delta_{i,i+\hat{\mu}}$.
We take for $d$-wave pairing as
\begin{equation}
\left\{
\begin{alignedat}{1}
~\Delta_{i,i+\hat{x}}= ~~ \Delta_s,\\
~\Delta_{i,i+\hat{y}}= -\Delta_s,
\end{alignedat}
\right.
\end{equation}
where $\Delta_s$ is a real constant representing the magnitude of the SC
order parameter.

In Fig. 4 we show the energy $E/N$ as a function of the electron number $N_e$
for several values of the superconducting gap function $\Delta_s$.
We used the periodic boundary condition in the $y$-th direction and antiperiodic
boundary condition in the $x$-th direction.
Since the electron number is not fixed in the BCS wave function, this applies
also to the correlated wave function $\psi_{\lambda-BCS}$. 
The number of electrons is adjusted by the chemical potential $\mu$.
We obtained the energy at $N_e=228$  through an extrapolation with respect to $N_e$.
To reduce the numerical error in the extrapolation, we employ the least squares
method.

We show the ground-state energy as a function of the SC order parameter $\Delta_s$
in Fig. 5 where $U=18t$ and $t'=0$ and calculations are carried out on
a $16\times 16$ lattice.  The energy $E/N$ has a minimum at a finite value of
$\Delta_s$.  This indicates a stability of the $d$-wave superconducting state.

\begin{figure}
\centering
\includegraphics[width=7.5cm]{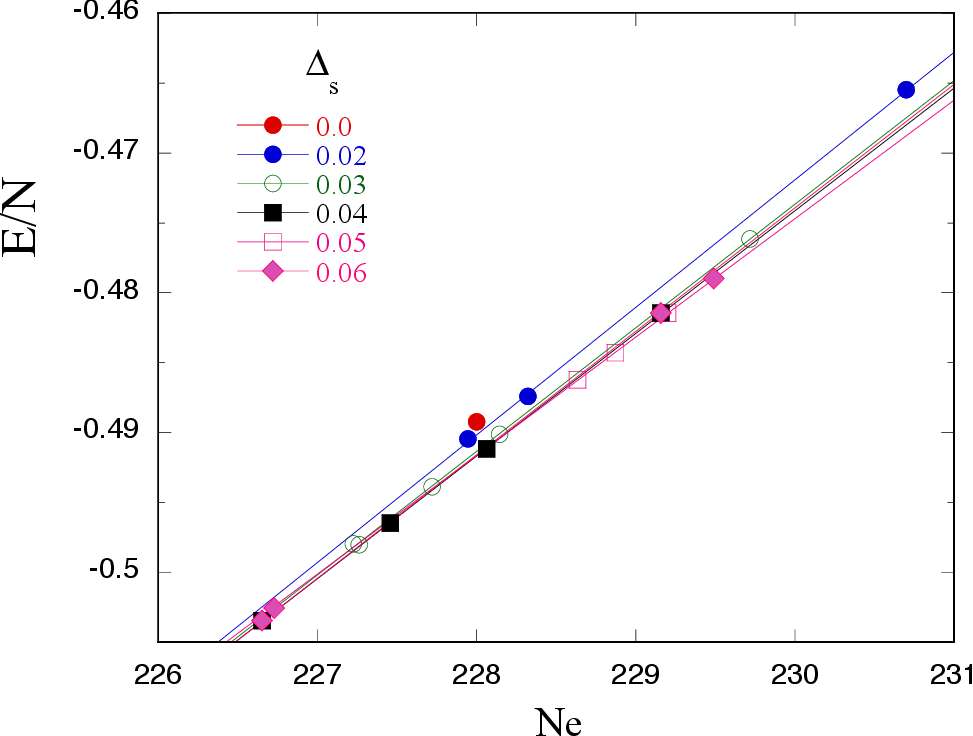}
\caption{(Color online)
The ground-state energy per site $E/N$ as a function of $N_e$ for
$\Delta_s= 0.02$, 0.03, 0.04, 0.06 and 0.08 for $U=18t$ and $t'=0$ on a
$16\times 16$ lattice.  The variational parameters are chosen as
$g= 0.005$ and $\lambda= 0.055$.
}
\label{fig4}
\end{figure}

\begin{figure}
\centering
\includegraphics[width=7.5cm]{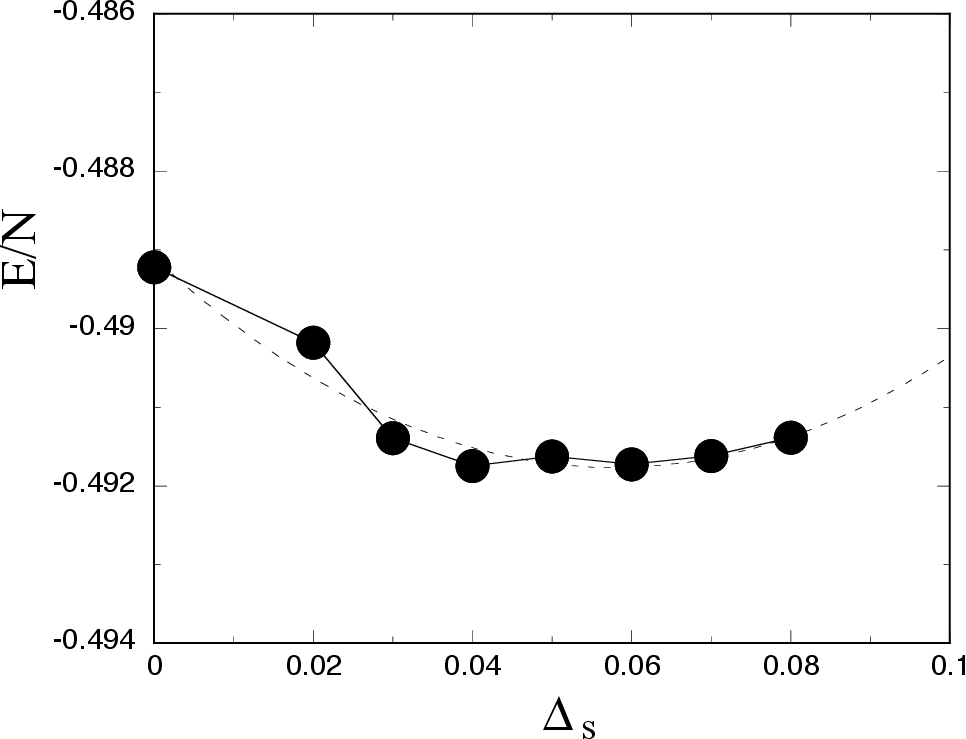}
\caption{
The ground-state energy per site as a function of the SC gap parameter $\Delta_s$
for $U=18t$ and $t'=0$ at $N_e=228$ on a $16\times 16$ lattice.
The variational parameters are the same as those in Fig. 4.
The Monte Carlo statistical errors are within the size of symbols.
}
\label{fig5}
\end{figure}

\section{Inhomogeneous states and superconductivity in the 2D Hubbard model}

We examine the stability of striped states employing the improved
wave function $\psi_{\lambda} = e^{-\lambda K}P_G\psi_{0}$.
Striped states are  stabilized in the framework of off-diagonal wave functions 
using $\psi_{\lambda}$.
As in previous studies, striped states are definitely stabilized when we include the
next nearest transfer $t'$ with negative values\cite{miy22}.

\subsection{Stripes in the case of $t'=-0.2$}

We consider the case with $t'=-0.2t$ in this subsection.
The strength of Coulomb repulsion is chosen as $U=18t$.
In this case the stripe state with the 8-lattice periodicity is stabilized near 1/8 doping.
We show the ground-state energy as a function of the AF order parameter $\Delta_{AF}$
for $\rho= 0.1, 0.2, 0.3, 0.4$ and 0.5 in Fig. 6 where the electron number is $N_e=224$ 
(1/8-doping) on the $16\times 16$ lattice.
The incommensurability is chosen as $\delta=1/4$ so that we have the 8-lattice
periodicity.
The charge density is oscillating in striped states, which is shown in Fig. 7
where the charge density is $n_i\equiv \langle n_{i\uparrow}+n_{i\downarrow}\rangle$
at the site ${\bf r}_i$. $n_i$ has a maximum at ${\bf r}={\bf r}_0$ where 
$({\bf r}_0)_x = 0$ (mod 4) and is uniform
in the y-direction.
The spin density $2S_z$ as a function of the distance is also shown in Fig. 8, 
where $2S_z=\langle n_{i\uparrow}-n_{i\downarrow}\rangle$.

\begin{figure}
\centering
\includegraphics[width=8.0cm]{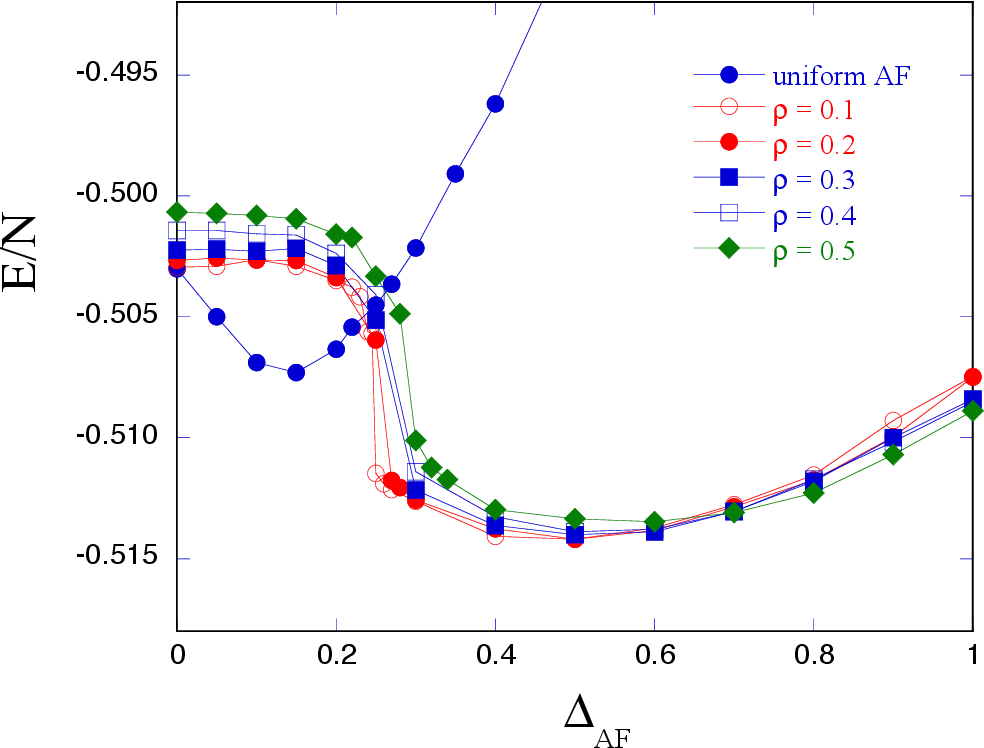}
\caption{(Color online)
The ground-state energy per site $E/N$ as a function of $\Delta_{AF}$ for
$U=18t$ and $t'=-0.2$ on the $16\times 16$ lattice.
The electron number is $N_e=224$.
We put $\rho = 0.1$, 0.2, 0.3, 0.4 and 0.5.  We included the energy of the
commensurate (uniform) AF state $\psi_{AF}$.
}
\label{fig6}
\end{figure}

\begin{figure}
\centering
\includegraphics[width=7.5cm]{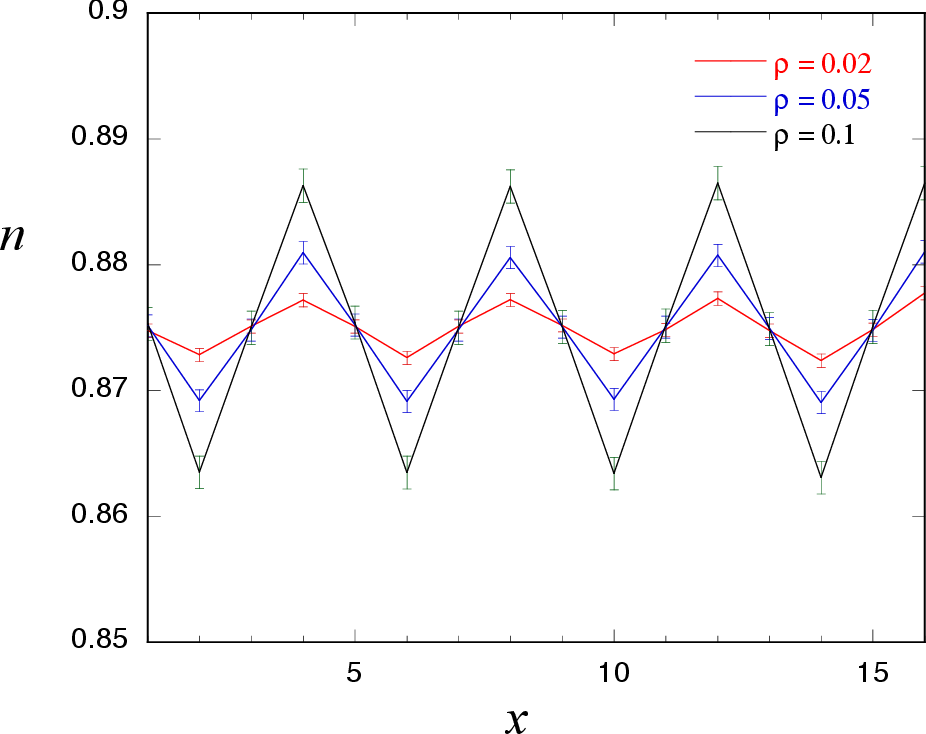}
\caption{(Color online)
The charge density $n=n_i$ as a function of the distance $x$ for $\rho = 0.02$, 0.05
and 0.1,  and $\Delta_{AF} = 0.4$ where stripes are along the $y$-direction.
The other parameters are the same as those in Fig. 6.
}
\label{fig7}
\end{figure}

\begin{figure}
\centering
\includegraphics[width=7.5cm]{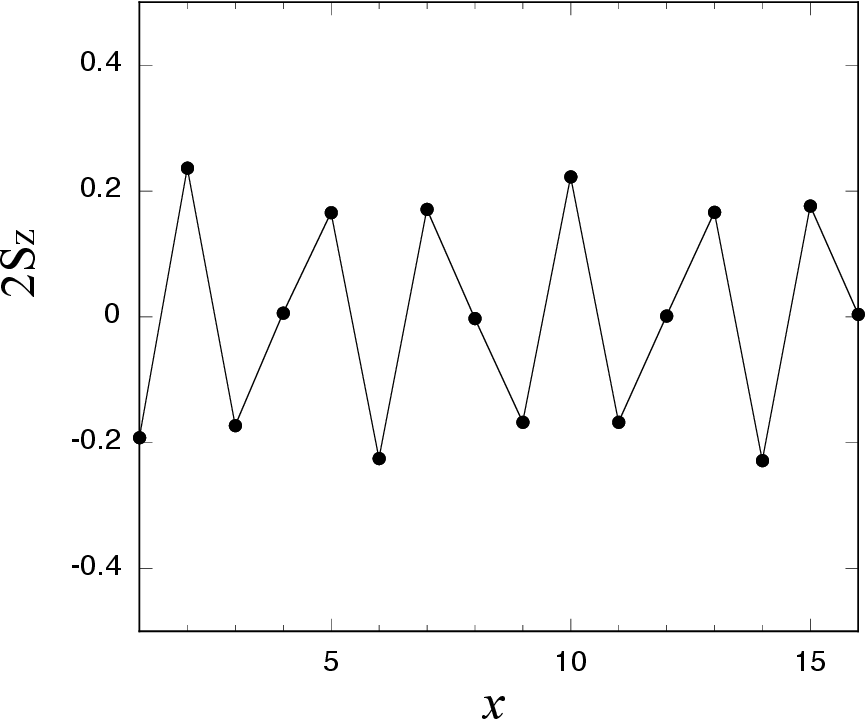}
\caption{
The spin density $2S_z$ as a function of the distance $x$ for $\rho = 0.1$ and
$\Delta_{AF}=0.4$.
The other parameters are the same as those in Fig. 6.
}
\label{fig8}
\end{figure}

\subsection{Charge-ordered nematic state in the case of $t'=0$}

In the case where $t'=0$, the striped state is not so easily stabilized
compared to the case with non-zero $t'$.
We show the energies of striped states as a function of $\rho$ for $U=18t$ in Fig. 9.
Here we considered the vertical and also diagonal striped states.
We have vanishing magnetic order parameter $\Delta_{AF}=0$ and non-zero charge
parameter $\rho$ at the minimum of the ground-state energy.
Thus the charge-ordered state without the AF ordering is realized in the ground state 
although the energy lowering is small.  This state may be called the nematic state.

\begin{figure}
\centering
\includegraphics[width=7.8cm]{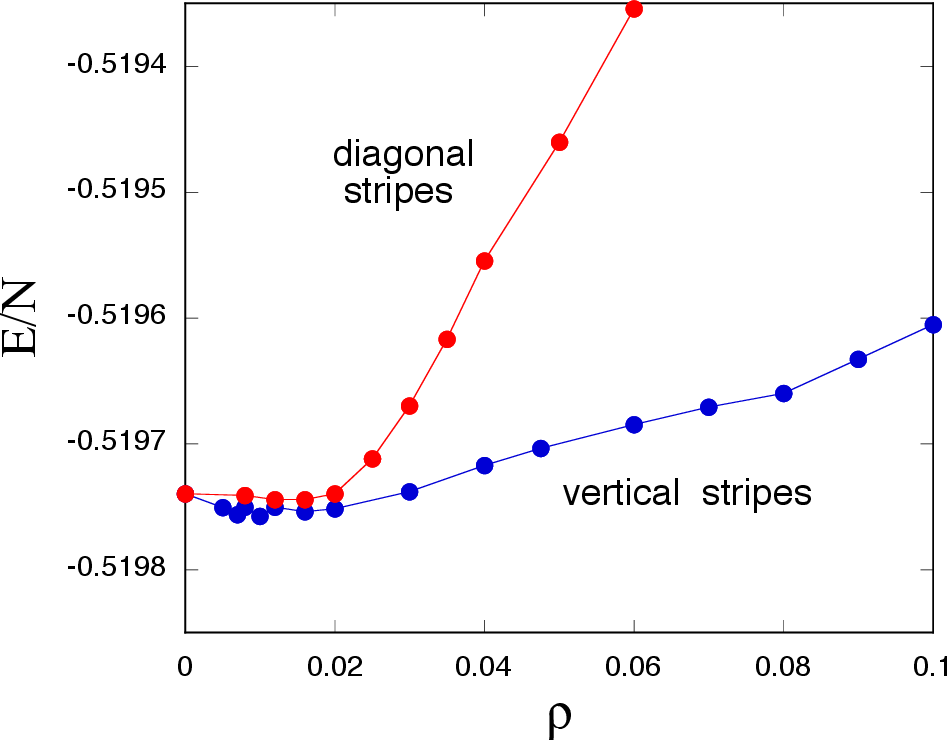}
\caption{(Color online)
The ground-state energy as a function of $\rho$ for $U=18t$ and $t'=0$
on the $16\times 16$ lattice with the electron number $N_e=224$. 
The AF parameter $\Delta_{AF}$ is set to 0.  We include the energy of the
diagonal stripe state as well as the vertical one with $\delta=1/4$.
}
\label{fig9}
\end{figure}

\subsection{Nematic superconductivity in the 2D Hubbard model}

Let us examine the correlated superconducting state in the charge inhomogeneous
state.  We consider the wave function $\psi_{\lambda-BCS}$ with the following
spatially oscillating gap function: 
\begin{equation}
\left\{
\begin{alignedat}{1}
~\Delta_{i,i+\hat{x}}=  \Delta_s\cdot (1+\alpha\cos({\bf Q}_c\cdot ({\bf r}_i-{\bf r}_0)+\gamma)),\\
~\Delta_{i,i+\hat{y}}= -\Delta_s\cdot (1+\alpha\cos({\bf Q}_c\cdot ({\bf r}_i-{\bf r}_0))),~~~
\end{alignedat}
\right.
\end{equation}
where $\alpha$ is a real variational parameter and we also introduce the phase $\gamma$ 
as a variational parameter. 
In the case with the charge order of 4-lattice periodicity, we use the following
gap function:
\begin{equation}
\left\{
\begin{alignedat}{1}
~\Delta_{i,i+\hat{x}}=  \Delta_s\cdot (1+\alpha\cos(\pi x/2-\pi/4)),\\
~\Delta_{i,i+\hat{y}}= -\Delta_s\cdot (1+\alpha\cos(\pi x/2)),~~~~~~~
\end{alignedat}
\right.
\end{equation}
where we put ${\bf r}_i=(x,y)$, and the hole (or electron) rich domains are on 
$x=4, 8, 12, 16$ on a $16\times 16$ lattice for $\alpha>0$ (or $\alpha<0$).  
This function is called the partially oscillating $d$-wave gap function in this paper.
We can also consider the other form for the gap function which is given by
\begin{equation}
\left\{
\begin{alignedat}{1}
~\Delta_{i,i+\hat{x}}=  \Delta_s\cdot \cos(\pi x/4+\pi/8),\\
~\Delta_{i,i+\hat{y}}= -\Delta_s\cdot \cos(\pi x/4).~~~~~~~
\end{alignedat}
\right.
\end{equation}
This is the gap function with full oscillation for which the amplitude has maxima on 
hole rich domains.

Let us consider the ground state for $U=18t$ and $t'=0$ with the number of electrons
given by $N_e=228$ on a $16\times 16$ lattice.
As is shown above, the charge-ordered nematic state is stable with $\rho=0.01$
in this case.
We show the ground-state energy $E/N$ as a function of the electron number $N_e$ in Fig. 10.
In Fig. 11 the $E/N$ is shown as a function of $\Delta_s$ for the uniform
$d$-wave pairing state and oscillating $d$-wave pairing.
The spatial dependence of the gap function $\Delta_{i,i+\hat{\mu}}$ is shown
in Fig. 12 where we set $\alpha=0.1$.
As is shown in Fig. 11, the partially oscillating $d$-wave pairing state has the
lowest energy and is most stable, and the energy of the uniform $d$-wave state is
close to that value. 
In Fig. 11 we take $\alpha=-0.1$ since the $d$-wave state with negative $\alpha$ has
lower energy than that with positive $\alpha$ such as $\alpha=0.1$.
In this case the gap function takes a maximum in electron rich domains.
This indicates that electrons forming Cooper pairs, not holes, give rise to superconductivity.
The result in Fig. 11 also shows that the pairing state with the fully oscillating gap 
function is unstable compared to that with the partially oscillating gap function.
In Fig. 13 we show the ground-state energy as a function of $\Delta_s$ for
several values of $\rho$.  The ground-state energy is the lowest at the finite value
of $\rho$ being given by $\rho\sim 0.03$.
This may confirm the coexistence of inhomogeneous charge order and superconductivity.

We give a comment on the phase separation in the state with weak inhomogeneity.
A phase-separated state is not stabilized since the energy of such a state is 
higher than that of the inhomogeneous state. 
This is because the kinetic energy gain is lost in a phase-separated state.
 
We here introduce the SC condensation energy as follows.
\begin{equation}
\Delta E = E(\Delta_s=0)-E(\Delta_{s,{\rm opt}}),
\end{equation}
where $E(\Delta_s)$ is the ground-state energy when the SC gap function is
$\Delta_s$, and $\Delta_{s,{\rm opt}}$ indicates the optimized value of $\Delta_s$.
We discuss the size dependence of $\Delta E$ for the charge-ordered state.
In Fig. 14, we show the SC condensation energy per particle $\Delta E/N_e$ as a function
of $1/N$, where we consider systems of $8\times 8$, $16\times 8$ and $16\times 16$
lattices since it is favorable that the length of each side is a multiple of 8. 
The result in Fig. 14 suggests that the superconducting state with this weak
inhomogeneity is stable in the thermodynamic limit.

We now discuss the size dependence of the inhomogeneous charge order.
The optimized value of $\rho$ is about 0.03 for both the $8\times 8$ and $16\times 16$
systems.  This suggests that the charge order still exists even in larger systems.
We estimated the energy gain due to the charge order where we define
\begin{equation}
\Delta E_{ch} = E(\Delta_{s,{\rm opt}}, \rho=0)-E(\Delta_{s,{\rm opt}}, \rho=\rho_{{\rm opt}}).
\end{equation}
Here $\rho_{{\rm opt}}$ stands for the optimized value of $\rho$.
We show $\Delta E_{ch}/N_e$ (the energy gain per particle) as a function of $1/N$ in Fig. 15.
Since the size of the lattices is limited, the extrapolation to the limit of large $N$ is not
necessarily conclusive. 
The result, however, suggests that $\Delta E_{ch}$ remains finite in the limit $N\rightarrow \infty$.

In Fig. 16, we show the charge density $n$ as a function of lattice sites for
$16\times 16$
and $8\times 8$ lattices, corresponding to those states in Figs. 14 and 15. 
Here we set $N_e=60$ for $8\times 8$ lattice and $N_e=228$ for
$16\times 16$ lattice, where electrons form the closed shell structure in both cases.
The charge density clearly shows an oscillating behavior in the real space and thus
the inhomogeneous charge-ordered state becomes stable in the ground state.

\begin{figure}
\centering
\includegraphics[width=7.5cm]{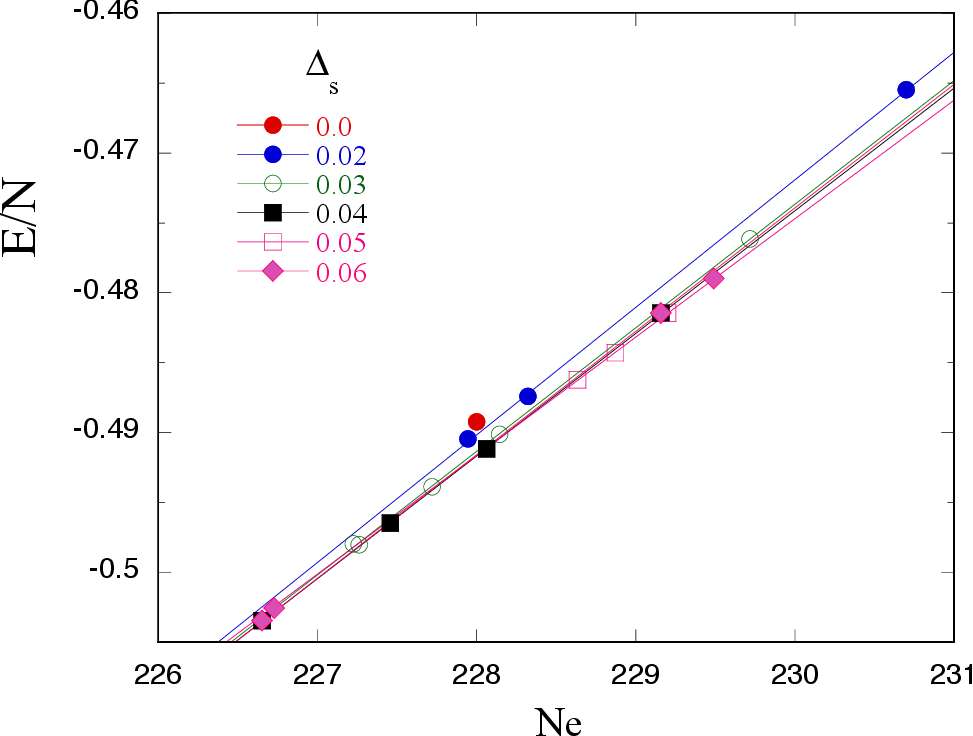}
\caption{(Color online)
The ground-state energy per site $E/N$ as a function of $N_e$ near 
$N_e=228$ for several values
of $\Delta_s$, where the parameters are $U=18t$ and $t'=0$ on a $16\times 16$
lattice. We used $g=0.005$, $\lambda=0.055$ and $\rho=0.01$. 
}
\label{fig10}
\end{figure}

\begin{figure}
\centering
\includegraphics[width=8.0cm]{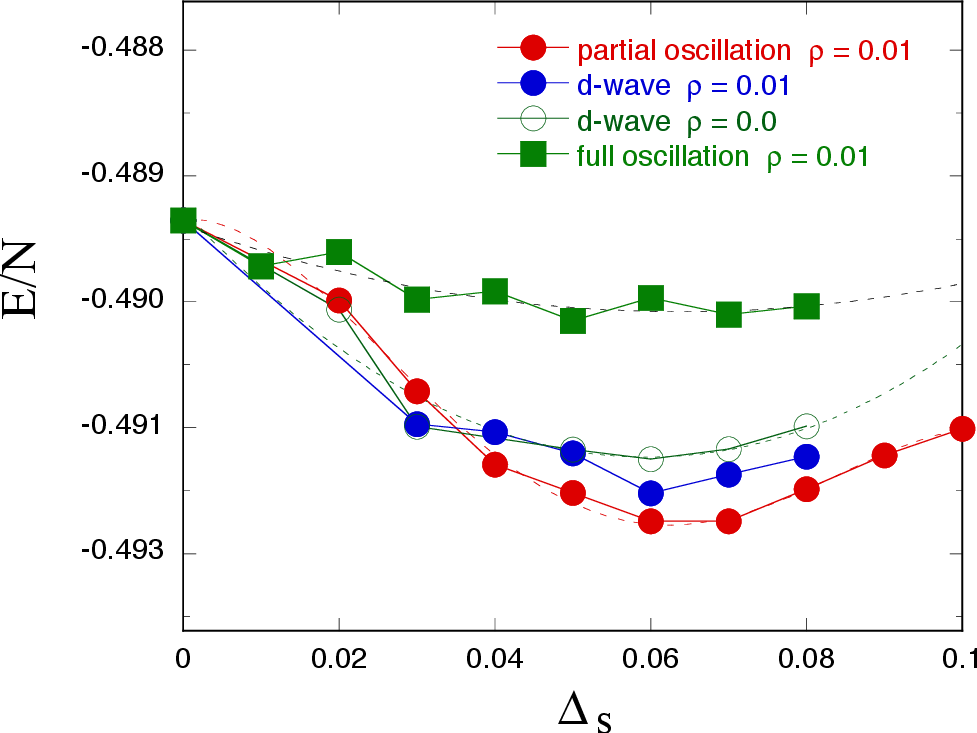}
\caption{(Color online)
The ground-state energy per site as a function of the SC gap parameter $\Delta_s$
for $U=18t$ and $t'=0$ at $N_e=228$ on a $16\times 16$ lattice.
The variational parameters are the same as those in Fig. 10.
We show ground-state energies for uniform $d$-wave symmetry, partially
oscillating $d$-wave ($\alpha=-0.1$) and fully oscillating $d$-wave pairing.
Monte Carlo statistical errors are within the size of symbols.
The dashed lines are guides to eyes.
}
\label{fig11}
\end{figure}

\begin{figure}
\centering
\includegraphics[width=5.5cm, angle=90]{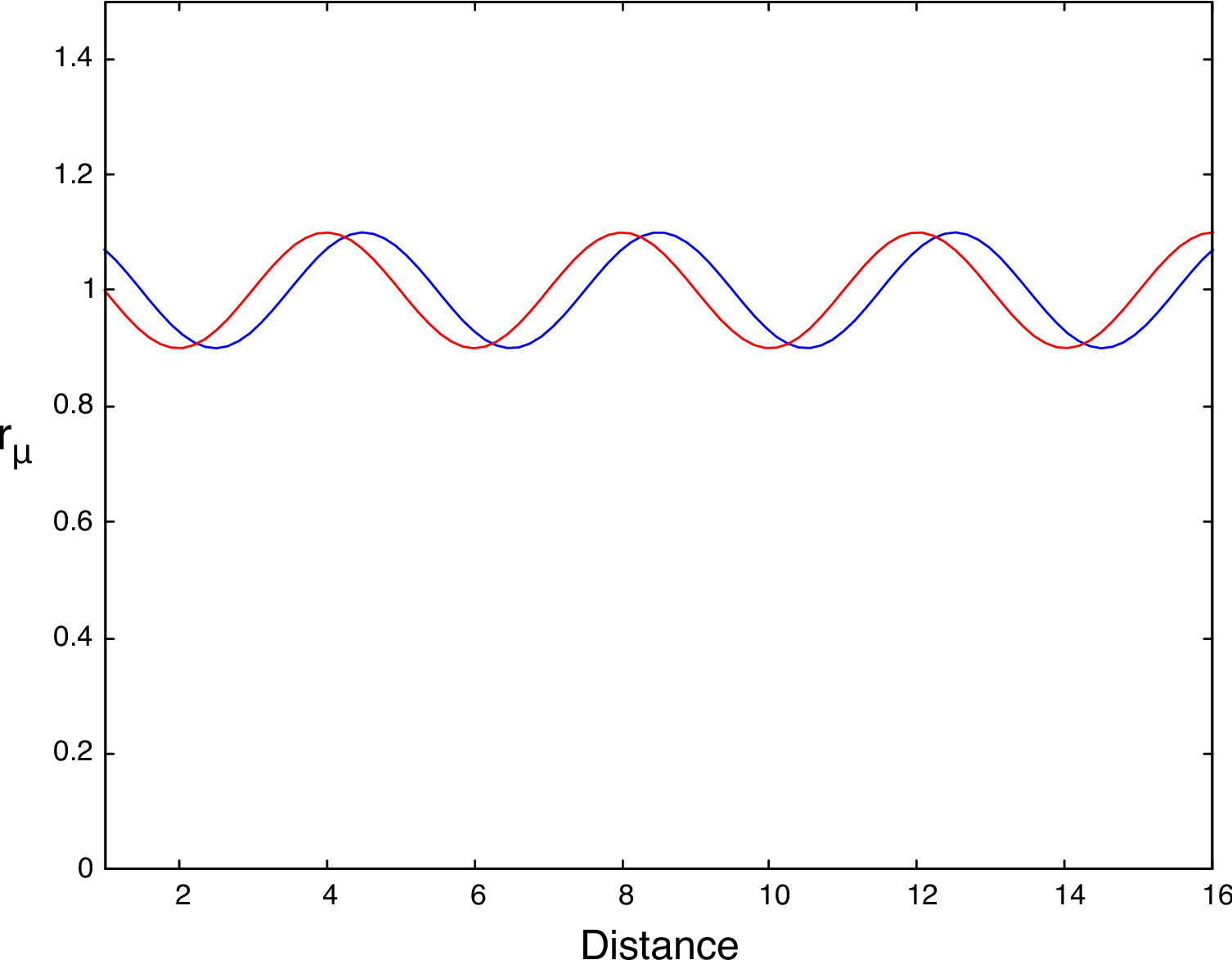}
\caption{(Color online)
The $x$ dependence of the gap function $r_{\mu}:=|\Delta_{i,i+\hat{\mu}}/\Delta_s|$
for $\mu=x, y$ and $\alpha=0.1$ (red for $\mu=y$ and blue for $\mu=x$).  
The phase $\pi/4$ is shifted for $\mu=x$ since the gap function is assigned at the
bond between two neighboring sites. 
}
\label{fig12}
\end{figure}

\begin{figure}
\centering
\includegraphics[width=7.5cm]{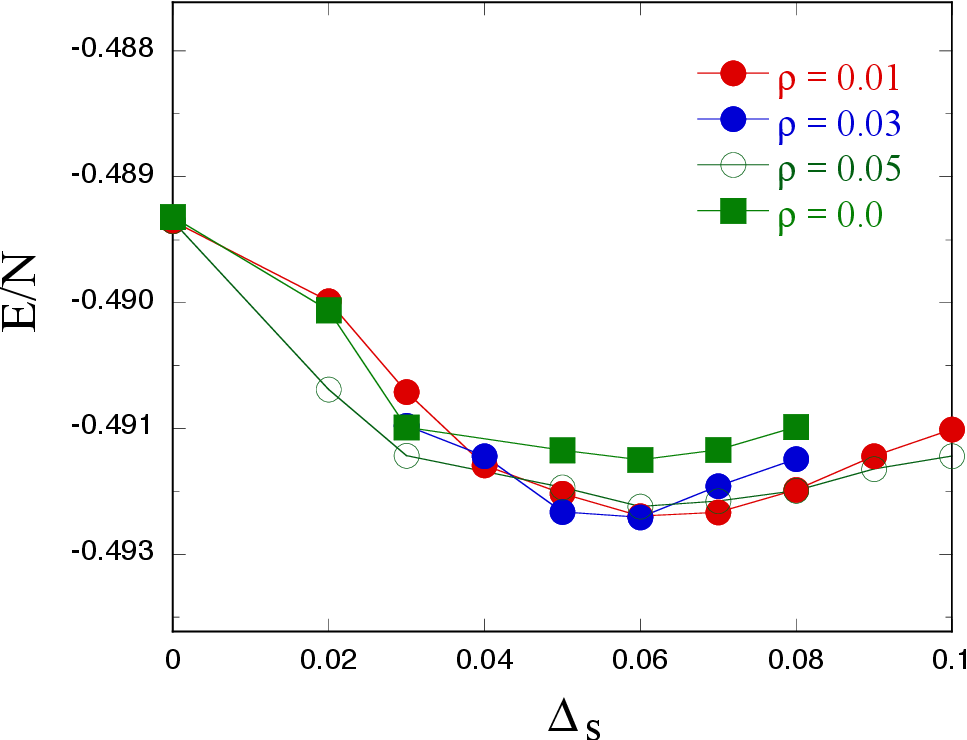}
\caption{(Color online)
The ground-state energy per site as a function of the SC gap parameter
$\Delta_s$ for $\rho=0.0$, 0.01, 0.03 and 0.05 on a $16\times 16$ lattice where
we used $U=18t$, $t'=0$ and $N_e=228$. The variational parameters are the same
as those in Fig. 11. 
Monte Carlo statistical errors are within the size of symbols. 
}
\label{fig13}
\end{figure}

\begin{figure}
\centering
\includegraphics[width=7.5cm]{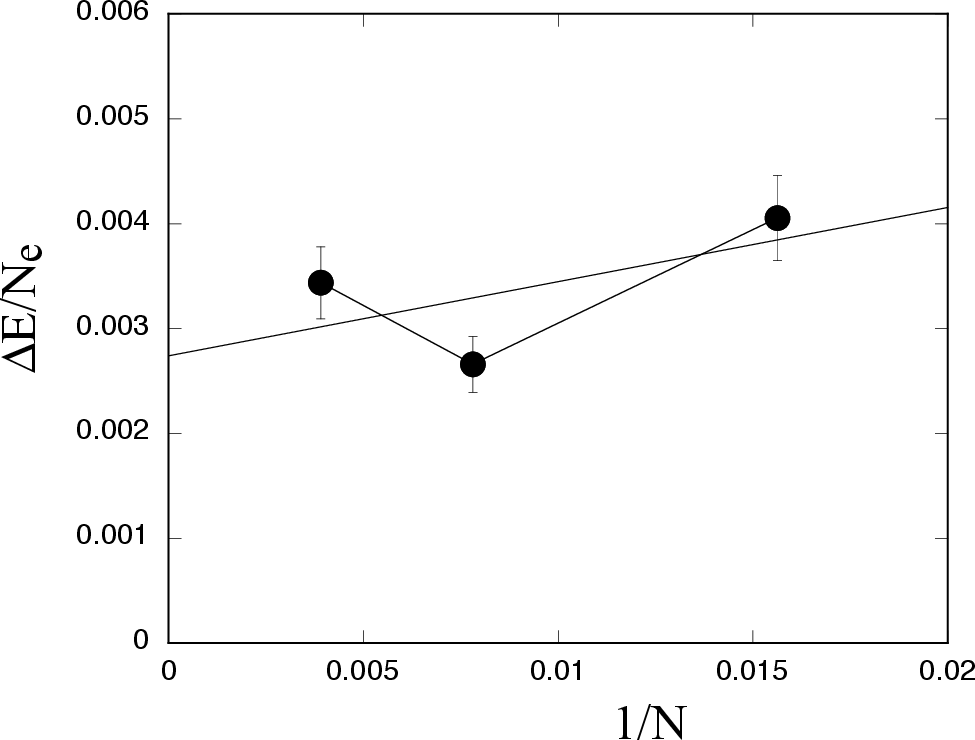}
\caption{(Color online)
The superconducting condensation energy per particle as a function of $1/N$ for $8\times 8$, 
$16\times 8$ and $16\times 16$ lattices.
The ground state is the $d$-wave pairing state coexisting with the weak charge order.
The straight line is given by the least squares method.
}
\label{fig14}
\end{figure}

\begin{figure}
\centering
\includegraphics[width=7.5cm]{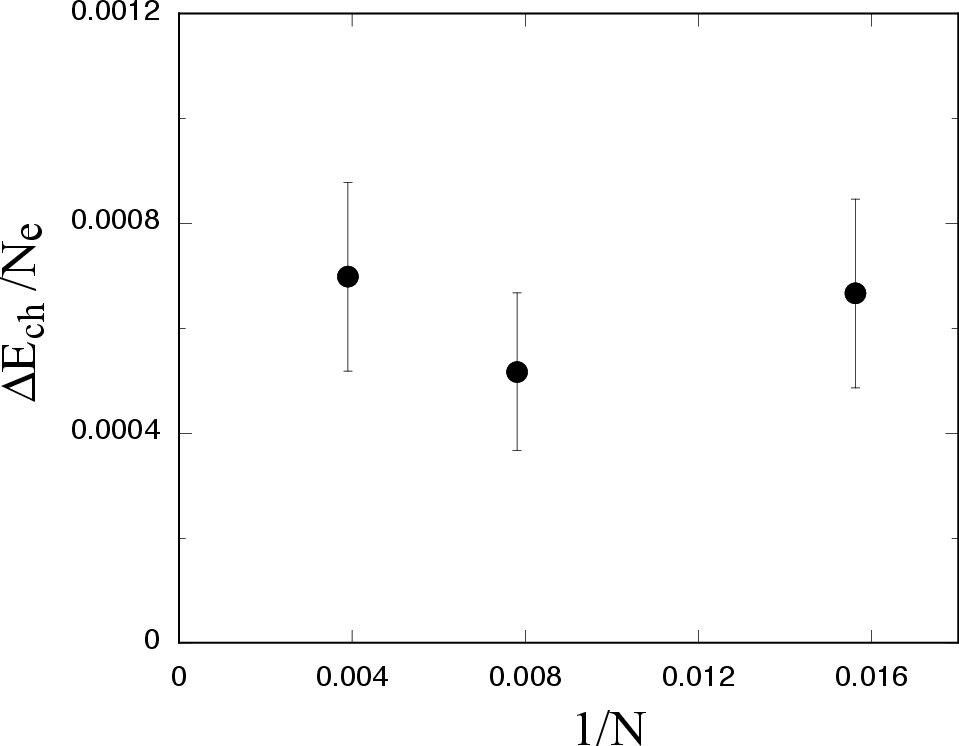}
\caption{(Color online)
The charge-energy gain per particle $\Delta E_{ch}/N_e$ as a function of $1/N$ for lattices 
$8\times 8$, $16\times 8$ and $16\times 16$ lattices.}
\label{fig15}
\end{figure}

\begin{figure}
\centering
\includegraphics[width=7.5cm]{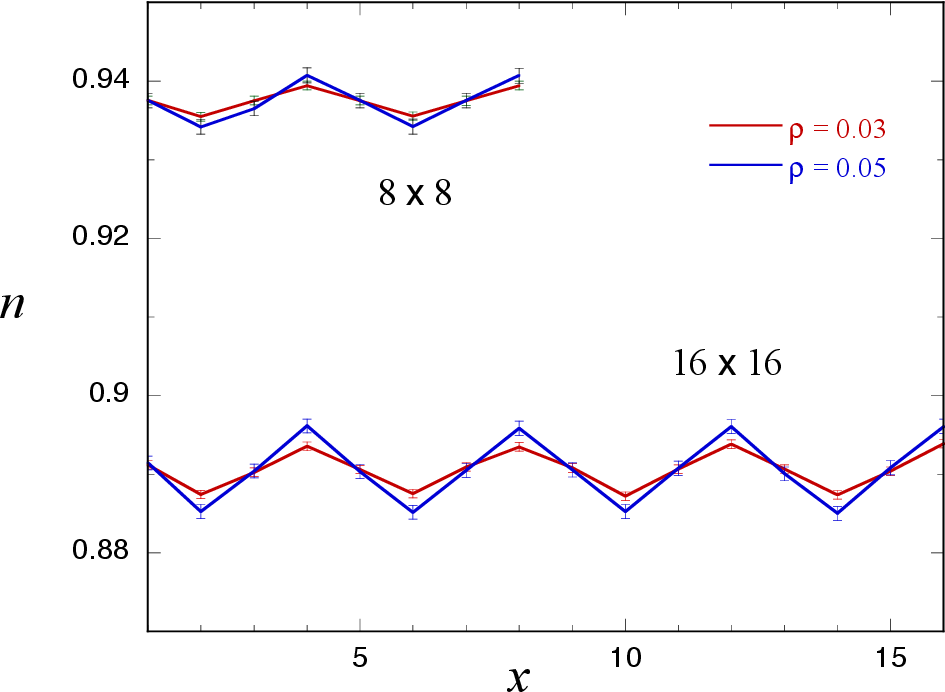}
\caption{(Color online)
The charge density $n=n_i$ as a function of the distance $x$ for $\rho=0.03$ and 0.05 on
$16\times 16$ and $8\times 8$ lattices, where $\Delta_{AF}=0$ and charge stripes are
along the $y$-direction.
The parameters are $U=18t$ and $t'=0$.
}
\label{fig16}
\end{figure}

\section{Pair correlation function in strongly correlated region}

In this section we investigate the behavior of the pair correlation function and 
the expectation value of the SC electron pairs $c_{i\uparrow}^{\dag}c_{j\downarrow}^{\dag}$.
We evaluate the SC correlation function
\begin{equation}
D_{sc}(\ell)= \langle \Delta^{\dag}(i)\Delta(i+\ell)\rangle,
\end{equation}
where the pair annihilation operator $\Delta(i)$ at the site $i$ is defined as
\begin{equation}
\Delta(i)= \Delta_x(i)+\Delta_{-x}(i)-\left( \Delta_y(i)+\Delta_{-y}(i)\right),
\end{equation}
with
\begin{equation}
\Delta_{\alpha}= c_{i\downarrow}c_{i+\hat{\alpha}\uparrow}
-c_{i\uparrow}c_{i+\hat{\alpha}\downarrow},
\end{equation}
for $\alpha=x$ and y.  $\hat{\alpha}$ indicates the unit vector in the $\alpha$-th
direction.

The expectation value of the SC order parameter is defined by
\begin{equation}
\Delta= \frac{1}{N}\sum_i \left( 
\langle c_{i\uparrow}^{\dag}c_{i+\hat{x}\downarrow}^{\dag} \rangle
-\langle c_{i\uparrow}^{\dag}c_{i+\hat{y}\downarrow}^{\dag} \rangle \right).
\end{equation}

We show the pair correlation function $D_{sc}(\ell)$ on a $10\times 10$ lattice
in Fig. 17 where the electron density is $n_e=0.88$ and we used $t'=0$.
This figure indicates that the long-range order of superconducting correlation
indeed exists and, however, the correlation function is reduced considerably
by strong electron correlation.
The Gutzwiller projection operator suppresses the pair correlation function as
shown in Fig. 17.
We exhibit $D_{sc}(\ell)$ on a $12\times 12$ lattice in Fig. 18;
$D_{sc}(\ell)$ shows a similar behavior as on $10\times 10$.
The strong electron correlation induces electron pairing and at the same time
the pair correlation function is reduced by electron correlation effect.
Thus, we might be able to say that the electron correlation is a doubled-edged sword
for superconductivity.

In Fig. 18 we introduced a new variational parameter $\Delta_{hs}$ to examine
a possibility to enhance the pair correlation $D_{sc}(\ell)$.
The parameter $\Delta_{hs}$ is introduced in the operator $K$ in the form
$\Delta_{hs}\sum_i\left( (c_{i\uparrow}^{\dag}c_{i+\hat{x}\downarrow}^{\dag}
+ {\rm h.c.})-(c_{i\uparrow}^{\dag}c_{i+\hat{y}\downarrow}^{\dag} + {\rm h.c.})\right)$.

The effect of the Gutzwiller operator on the superconducting order parameter 
(gap function)
$\Delta$ is shown in Fig. 19 where the expectation value $\Delta$ of the order
parameter is shown as a function of $1-g$.
$\Delta$ decreases as $g$ decreases from 1 to 0.  This indicates that the expectation
value of the order
parameter $\Delta$ is reduced by strong electron correlation\cite{yan24}.   
It remains, however, finite as far as $g$ is finite $g>0$.
When $g$ is fixed, the estimated SC order parameter in the $16\times 16$ lattice
is greater than that in the $10\times 10$ lattice. 
This shows that the SC order parameter does not necessarily decrease as the
system size increases.

In Fig. 20, the gap function $\Delta$ is shown as a function of the input variational parameter
$\Delta_s$ on a $16\times 16$ lattice for $U=18t$ and $N_e=228$.  
$\Delta$ is reduced compared to the variational parameter $\Delta_s$ when $\Delta_{hs}=0$, 
which is similar to the behavior of the pair correlation function $D_{sc}(\ell)$.   
When $\Delta_{hs}$ is introduced, $\Delta$ becomes large due to the effect of
correlation operator $e^{-\lambda K}$.
The actual value of $\Delta_{hs}$ after the optimization is, however, small compared to
$\Delta_s$ for $U=18t$.  
In the inhomogeneous charge-ordered state with $\rho>0$, $\Delta$ remains the same
and $\Delta$ is almost independent of the parameter $\rho$.
We discuss here the size dependence of the gap function and pair correlation
function.  As shown in Fig. 19, the size dependence of $\Delta$ is not large, and 
when the Gutzwiller parameter $g$ is the same, there is the tendency that $\Delta$
increases as the system size increases. 
The pair correlation functions at large distance on $10\times 10$ and $12\times 12$
are almost the same and their size dependence is very small. 
It follows from these results that superconductivity does not vanish in the limit of
large $N$.

In order to obtain an enhanced pair correlation function, we must choose a smaller
value of $U$ in calculations.  
We show $D_{sc}(\ell)$ for $U=12t$ and $U=10t$ in Fig. 21 on a $10\times 10$ lattice. 
The pair correlation functions for $U=10t$ and $12t$ are larger than that for $U=18t$
for any value of $\ell$.
We mainly used $U=18t$ in this paper because the AF order vanishes when
$U$ is as large as $U\simeq 18t$ and the pure $d$-wave SC state becomes stable.
When the wave function is improved by multiplying by correlation operators further, 
the AF correlation is reduced and we could choose a smaller value of $U$ for the
realization of $d$-wave pairing state.

\begin{figure}
\centering
\includegraphics[width=7.5cm]{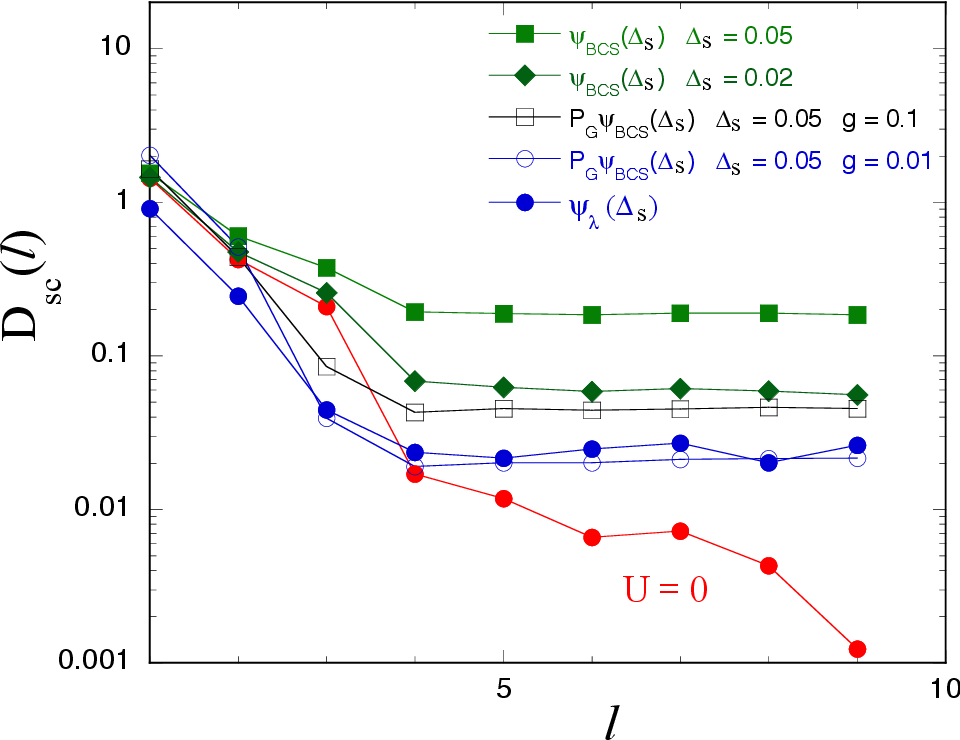}
\caption{(Color online)
The pair correlation function $D_{sc}(\ell)$ as a function of the lattice site $\ell$ 
on a $10\times 10$
lattice with $U=18t$, $t'=0$ and $N_e=88$.
The lattice sites $\ell$ are $\ell$=(1,1), (1,2), (1,3), (1,4), (1,5), (2,5), (3,5),
(4,5) and (5,5) and the site $i$ is chosen as $i=(1,1)$.
The figure includes $D_{sc}(\ell)$ for $U=0$, the non-interacting BCS wave function
with the gap function $\Delta_s=0.05$ and 0.02 and the the Gutzwiller projected
BCS function with $\Delta_s=0.05$.
The result for $\psi_{\lambda}$ indicates $D_{sc}(\ell)$ for the optimized
wave function with $\Delta_s=0.05$.
}
\label{fig17}
\end{figure}

\begin{figure}
\centering
\includegraphics[width=7.5cm]{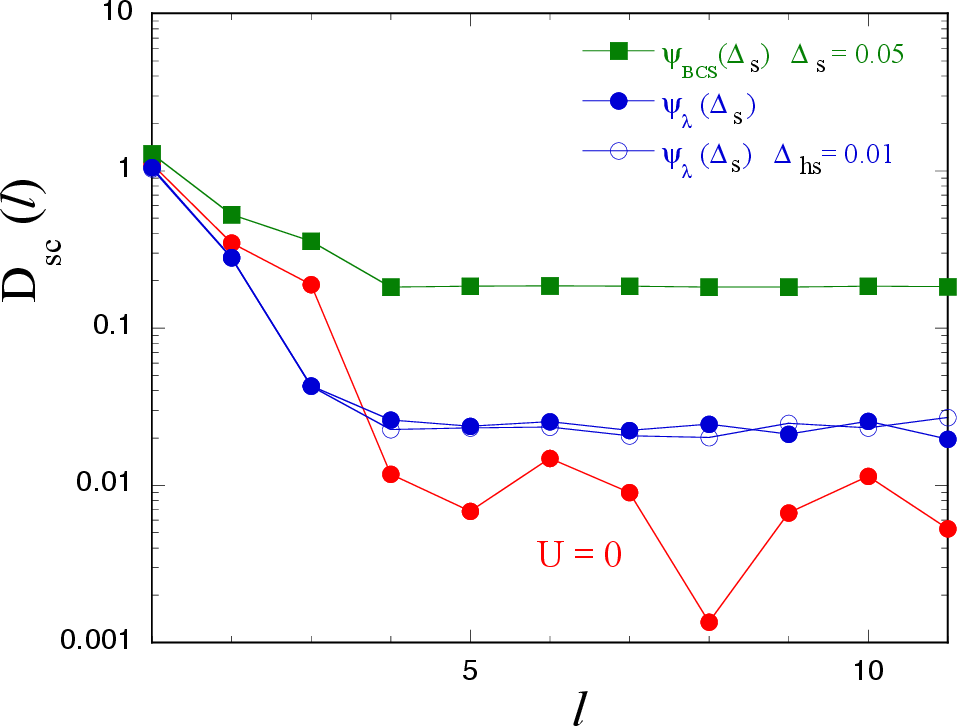}
\caption{(Color online)
The pair correlation function $D_{Sc}(\ell)$ as a function of the lattice site $\ell$
on a $12\times 12$ lattice.
The lattice sites $\ell$ are $\ell$=(1,1), (1,2), (1,3), (1,4), (1,5), (1,6), (2,6),
(3,6), (4,6), (5,6) and (6,6) for the site $i=(1,1)$.
We show $D_{sc}(\ell)$ for the non-interaction case ($U=0$, red circles), the non-interacting
BCS function with $\Delta_s=0.05$ and $\psi_{\lambda}$ with $\Delta_s=0.05$.
We also include the result for $\psi_{\lambda}$ with an additional variational
parameter $\Delta_{hs}$, where $\Delta_{hs}$ is the parameter introduced in the
kinetic term $K$.
}
\label{fig18}
\end{figure}

\begin{figure}
\centering
\includegraphics[width=7.0cm]{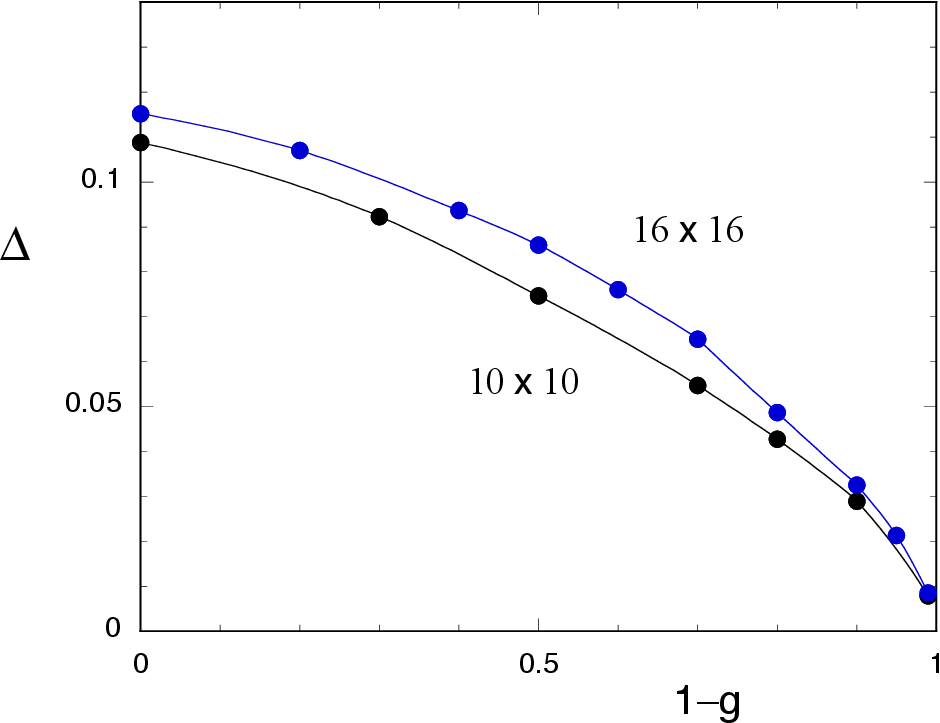}
\caption{(Color online)
The expectation value of SC order parameter $\Delta$ as a function of the Gutzwiller
parameter $1-g$ for the BCS-Gutzwiller function $P_G\psi_{BCS}$ on $10\times 10$
and $16\times 16$ lattices.  We set $N_e=88$ on $10\times 10$ and $N_e=228$ on
$16\times 16$ lattice.
}
\label{fig19}
\end{figure}

\begin{figure}
\centering
\includegraphics[width=7.5cm]{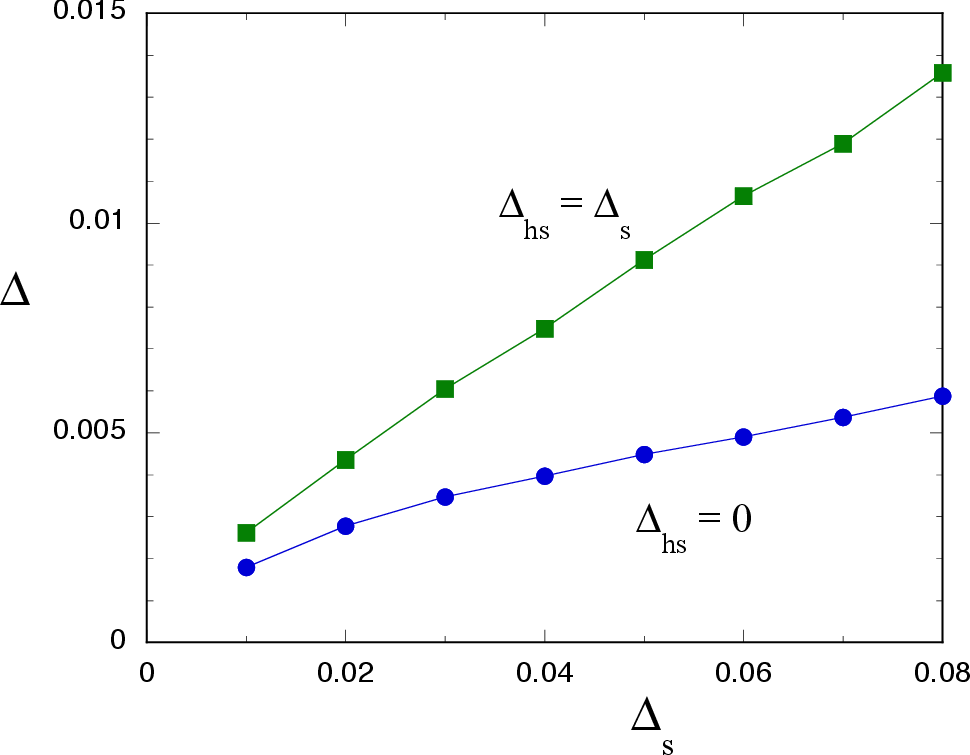}
\caption{(Color online)
The expectation value of SC order parameter $\Delta$ as a function of the variational
parameter $\Delta_s$ for the optimized wave function $\psi_{\lambda}$ on a $16\times 16$ 
lattice for $U=18t$, $t'=0$ and $N_e=228$.  We also show the result of $\Delta$ with 
the variational parameter
$\Delta_{hs}=\Delta_s$.
The expectation value of the SC order parameter is reduced due to the electron
correlation effect. 
}
\label{fig20}
\end{figure}

\begin{figure}
\centering
\includegraphics[width=7.5cm]{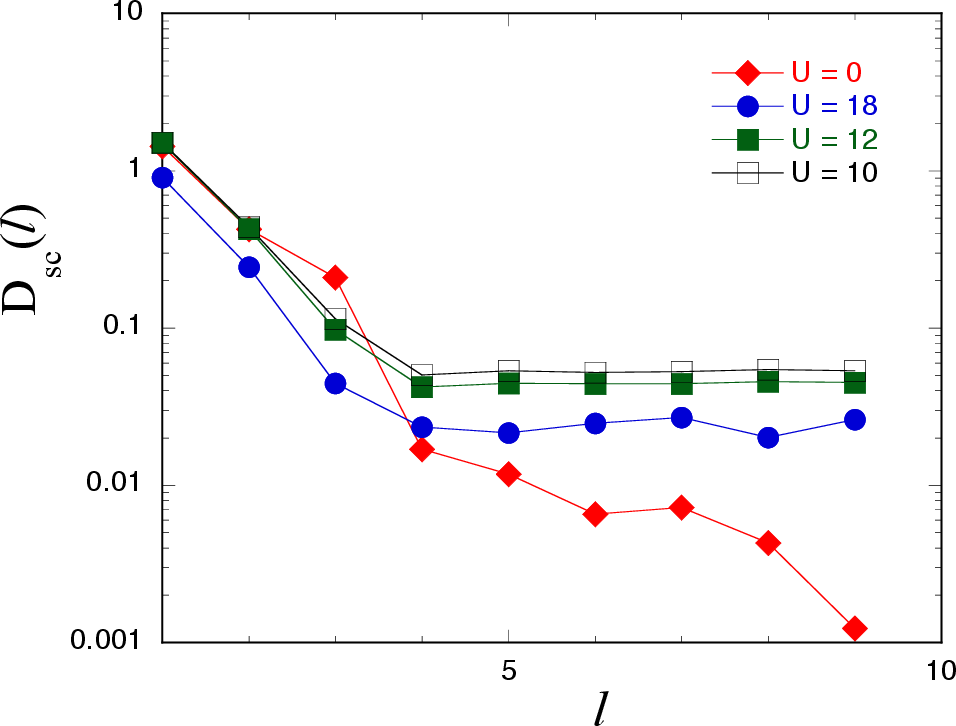}
\caption{(Color online)
The pair correlation function $D_{sc}(\ell)$ as a function of the lattice site $\ell$ 
on a $10\times 10$ lattice for $U=18t$, $U=12t$ and $U=10t$ with $t'=0$ and
$N_e=88$.
The lattice sites $\ell$ are $\ell$=(1,1), (1,2), (1,3), (1,4), (1,5), (1,6), (2,6),
(3,6), (4,6), (5,6) and (6,6) for the site $i=(1,1)$.
For the moderate value of $U$ in the strongly correlated region the pair correlation
$D_{sc}(\ell)$ is enhanced compared to that for $U=18t$.
}
\label{fig21}
\end{figure}

\section{Summary}

We have investigated the ground-state properties in the strongly correlated region of
the two-dimensional Hubbard model by using the optimization variational Monte
Carlo method.  We examined the inhomogeneous ground state at 1/8 hole doping 
near the optimal doping region.  The strength of the on-site Coulomb repulsive interaction
is chosen as $U/t=18$ which is much larger than the bandwidth.
We examined the case of this value in this paper because the antiferromagnetic 
correlation is extremely
suppressed when $U$ is as large as $18t$.
The striped state is in general stable when the doping rate $x$ is near $x=1/8$.
Both the spin and charge inhomogeneous distributions are cooperated to lower the
ground-state energy in stripe states.  This mechanism well works for the negative $t'$.
In the case of vanishing $t'$, stripes are not stabilized when $U/t$ is large 
because the antiferromagnetic state becomes unstable and its correlation is weak
for $t'=0$.  Thus for $t'=0$ a state with charge inhomogeneity is realized without
magnetic ordering.

The stability of charge inhomogeneous state may be closely related to the kinetic
energy effect in 
superconductivity\cite{miy22,mai04,oga06,gul12,toc16,fen03,wro03,guo07,yan21,yan21b}.
The kinetic energy gain along one-dimensional hole domains stabilizes the 
charge-ordered state.  It has been argued that the kinetic energy effect becomes important
in the strongly correlated region\cite{yan21b}.  We expect that this effect also plays an 
essential role in the emergence of charge inhomogeneity.   

Our question is how superconducting state is realized in this inhomogeneous state.
We have shown that superconductivity coexists with inhomogeneous charge ordering and
further that the superconducting condensation energy increases in the charge-ordered
state.  The superconducting gap function has a spatial dependence in accordance with that
of charge density.

We calculated the pair correlation function $D_{sc}(\ell)$ as a function of lattice sites.  
This indicates that the long-range order exists in the wave function adopted in
this paper.  The effect of strong electron correlation on superconductivity, however, 
is not trivial.  The $d$-wave superconductivity is induced by electron correlation 
and at the same time the pair correlation function is reduced by the electron 
correlation.  This is the reason why it is not easy to observe a clear evidence of
enhanced pair correlation function in the 2D Hubbard model. 
The superconducting gap function $\Delta$, which is finite in our
wave function, also shows a similar behavior.  These results indicate that the 
moderate value of $U$ being as large as the bandwidth might be better for the
realization of high-temperature superconductivity.

\section{Acknowledgment}

The author expresses his sincere thanks to K. Yamaji and M. Miyazaki for
valuable discussions.

A part of computations was supported by the Supercomputer
Center of the Institute for Solid State Physics, the University of
Tokyo. Numerical calculations were carried out also on the Supercomputer system 
Yukawa-21 of the Yukawa Institute
for Theoretical Physics, Kyoto University.

*t-yanagisawa@aist.go.jp

\end{document}